\documentclass[iop,apjl]{emulateapj}
\shortauthors{R. FUSCO-FEMIANO ET AL.}
\shorttitle{SZ VS. X-RAY VIEWS OF COMA}

\slugcomment{Accepted by ApJL}

\begin{document}

\title{The \textsl{Planck} Sunyaev-Zel'dovich vs. the X-ray View of the Coma Cluster}
\author{R. Fusco-Femiano$^{1}$, A. Lapi$^{2,3}$, A. Cavaliere$^{2,4}$}
\affil{$^1$IAPS-INAF, Via Fosso del Cavaliere, 00133 Roma,
Italy.}\affil{$^2$Dip. Fisica, Univ. `Tor Vergata', Via Ricerca Scientifica
1, 00133 Roma, Italy.}\affil{$^3$SISSA, Via Bonomea 265, 34136 Trieste,
Italy.}\affil{$^4$INAF, Osservatorio Astronomico di Roma, via Frascati 33,
00040 Monteporzio, Italy.}

\begin{abstract}
The \textsl{Planck} collaboration has recently published precise and resolved
measurements of the Sunyaev-Zel'dovich effect in Abell 1656 (the Coma cluster
of galaxies), so directly gauging the electron pressure profile in the
intracluster plasma. On the other hand, such a quantity may be also derived
from combining the density and temperature provided by X-ray observations of
the thermal bremsstrahlung radiation emitted by the plasma. We find a
model-independent tension between the SZ and the X-ray pressure, with the SZ
one being definitely lower by $15-20\%$. We propose that such a challenging
tension can be resolved in terms of an additional, non-thermal support to the
gravitational equilibrium of the intracluster plasma. This can be
straightforwardly included in our Supermodel, so as to fit in detail the
\textsl{Planck} SZ profile while being consistent with the X-ray observables.
Possible origins of the nonthermal component include cosmic-ray protons,
ongoing turbulence, and relativistic electrons; given the existing
observational constraints on the first two options, here we focus on the
third. For this to be effective, we find that the electron population must
include not only an energetic tail accelerated to $\gamma\ga 10^3$
responsible for the Coma radiohalo, but also many more, lower energy
electrons. The electron acceleration is to be started by merging events
similar to those which provided the very high central entropy of the thermal
intracluster plasma in Coma.
\end{abstract}


\section{Introduction}

We are motivated by the recent, spatially resolved measurements with the
\textsl{Planck} satellite by Ade et al. (2012) of the Sunyaev-Zel'dovich
(1980; SZ) effect in Abell 1656, the very rich, nearby cluster in Coma
Berenices at $z=0.023$.

The thermal SZ effect describes how the temperature of crossing CMB photons
is modulated by the Compton upscattering off the hot electrons in the
intracluster plasma (ICP). Its strength is given by the Comptonization
parameter $y\equiv (\sigma_T/m_e c^2)\, \int{\mathrm{d}\ell}\, p_e(r)$
integrated along line-of-sights across the cluster. It directly probes the
electron thermal pressure $p_e\approx p\, (2+2 X)/(3+5 X)\geq 0.5\, p$, here
written in terms of the ICP pressure $p$; with the cosmic hydrogen abundance
$X\approx 0.76$, their ratio reads $p_e/p = 0.52$. Compared with previous
observations including \textsl{WMAP}'s (see Komatsu et al. 2011, and
references therein), the resolved \textsl{Planck} data improve the SZ probing
of the cluster core and extend it into the outskirts, providing a handle to
the complex astrophysical processes in the ICP to be discussed here.

\begin{figure*}
\begin{center}
\epsscale{1.15}\plottwo{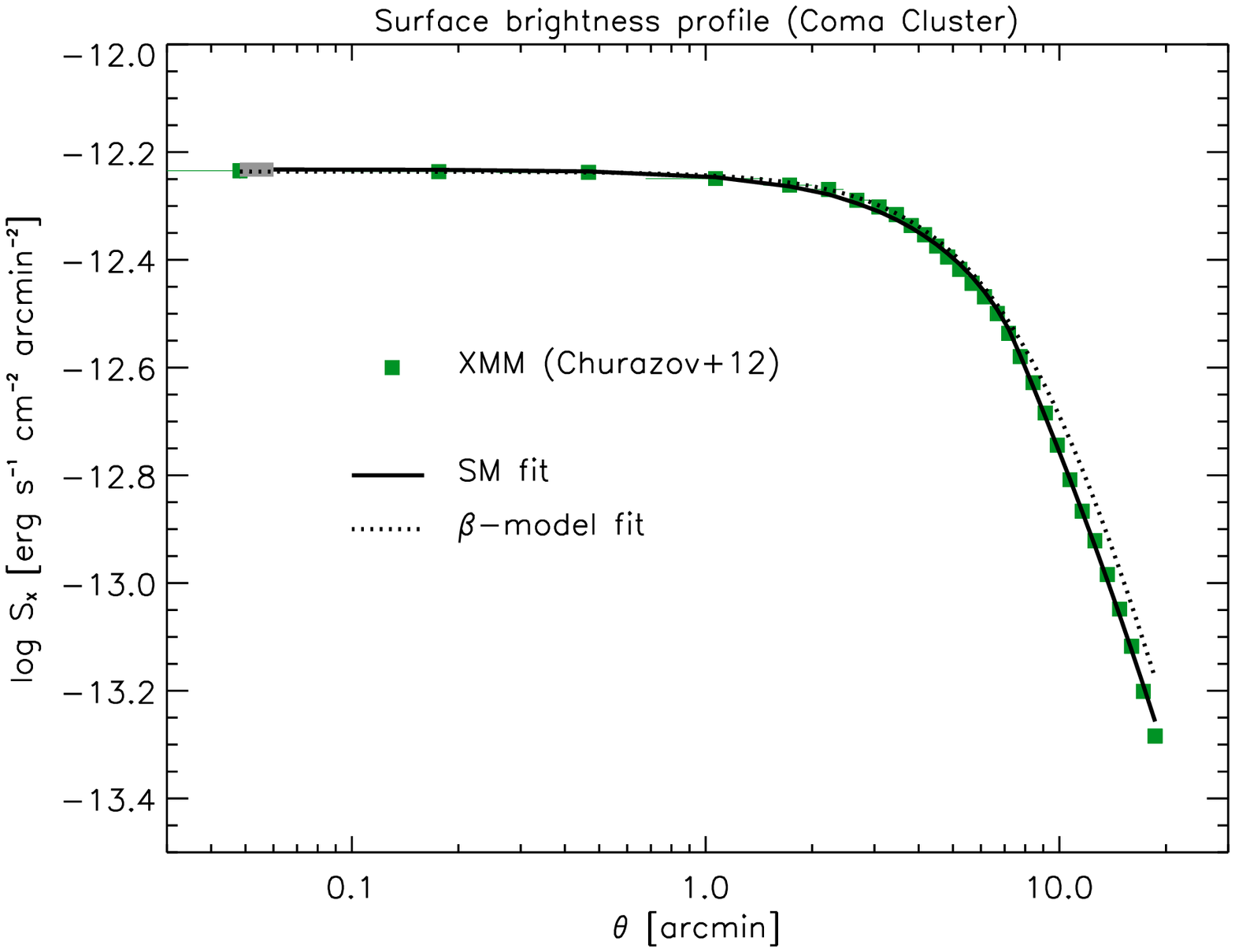}{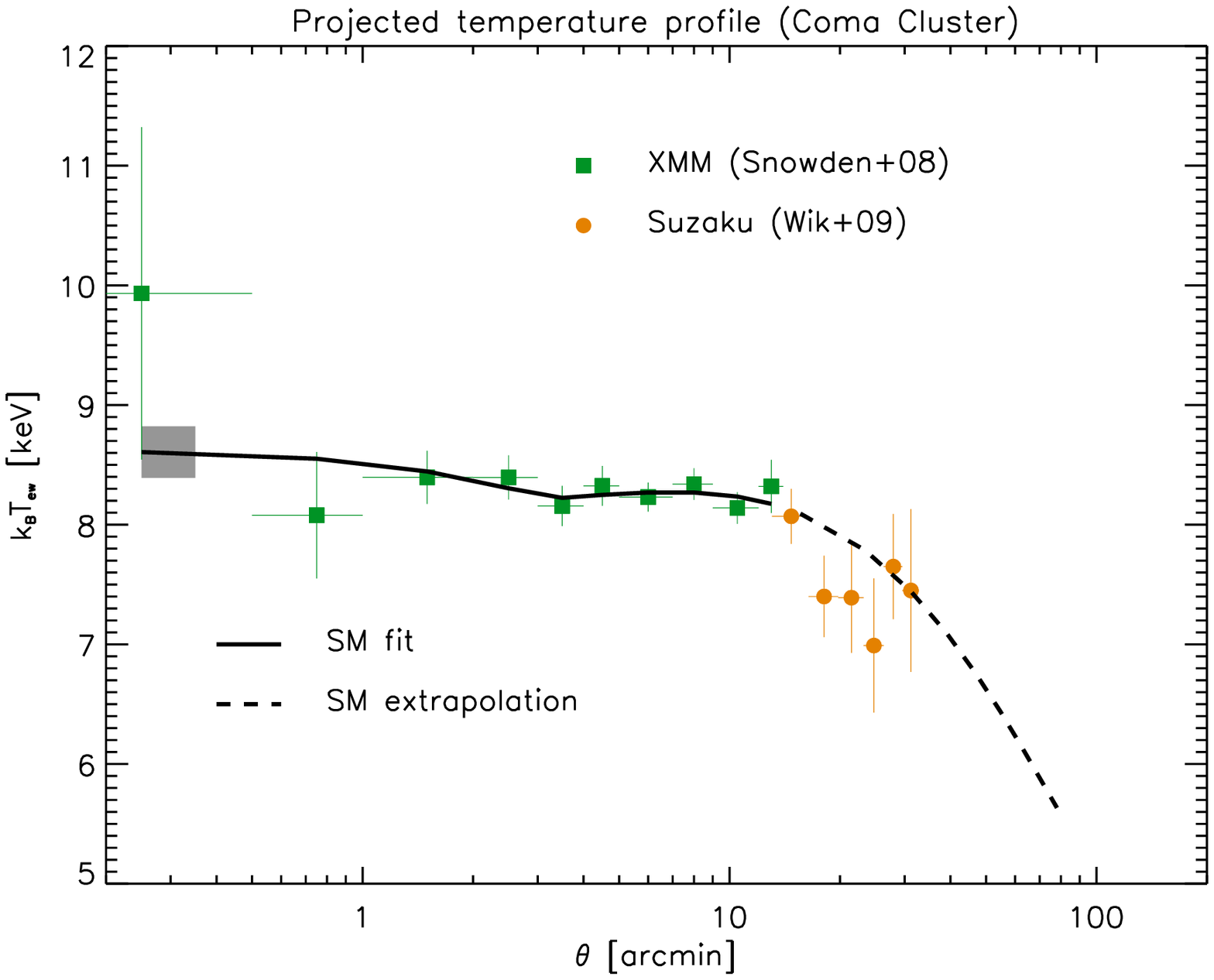}\caption{Top panel: Profile of the
X-ray surface brightness in the Coma Cluster; green squares refer to the
\textsl{XMM-Newton} data by Churazov et al. (2012), dotted line shows their
$\beta-$model fit, while solid line illustrates our SM outcome. Bottom panel:
Projected profile of the emission-weighted temperature in the Coma Cluster;
green squares refer to the \textsl{XMM-Newton} data by Snowden et al. (2008),
orange circles to the \textsl{Suzaku} data by Wik et al. (2009), solid line
is our SM outcome (obtained on fitting the \textsl{XMM-Newton} data), with
the dashed line representing its extrapolation into the outskirts out to the
virial radius $R=2.2$ Mpc (see Churazov et al. 2012). In both panels the
shaded areas show the associated $2-\sigma$ uncertainties.}
\end{center}
\end{figure*}

On the other hand, the ICP pressure $p\approx n k_B T/\mu$ (with the mean
molecular weight $\mu\approx 0.60$) may be also derived from combining the
density $n$ and temperature $T$ provided by X-ray observations of the thermal
bremsstrahlung radiation emitted by the plasma. The overall trend emerging
from the \textsl{Planck} data is toward a \emph{deficit} in the SZ relative
to the X-ray pressure, see Ade et al. (2012). In detail, to fit the SZ data
these authors based on empirical formulae suggested by numerical simulations
(see Nagai et al. 2007) and by X-ray analyses (see Arnaud et al. 2010). These
formulae provide a `universal' pressure profile for the whole cluster
population, or a specific version for unrelaxed clusters. However, when
applied to fit the precise \textsl{Planck} SZ data these formulae perform
inadequately, as discussed by Ade et al. (2012); specifically, the first
version turns out to overshoot the data in the core, and both to appreciably
undershoot them in the outskirts, well beyond the quoted uncertainties. Aimed
modifications of the parameter values in the fitting formulae, that include
suppression of unphysical central divergencies, can improve the SZ fits at
the cost of inconsistencies with the X-ray pressure. As discussed by the
above authors, this is also the case with multiparametric fitting formulae of
the type proposed by Vikhlinin et al. (2006) for the X-ray observables.

Thus we are stimulated to use an \textit{orthogonal} approach, provided by
the Supermodel (SM; Cavaliere et al. 2009); this yields a direct
\textit{link} between the X-ray and the SZ observables (see Lapi et al.
2012). We specify below how the SM is based on a physically motivated run of
the entropy $k(r)\equiv k_B T(r)/n^{2/3}(r)$ that underlies the ICP support
in the gravitational potential well provided by the cluster dark matter, to
approach hydrostatic equilibrium.

The SM improves in a number of respects upon the classic isothermal
$\beta-$model (Cavaliere \& Fusco-Femiano 1976), adopted by Mohr et al.
(1999) and Churazov et al. (2012) to closely describe the central X-ray
brightness profile in Coma. In fact, the SM incorporates updated, weakly
cusped distributions of the dark matter (see Lapi \& Cavaliere 2009). It also
accurately describes the central conditions both in cool, and in non-cool
core clusters as Coma (see Molendi \& Pizzolato 2001). Finally, it also
describes the region beyond a few $10^2$ kpc where $T$ declines outwards
while $n$ drops, so that the entropy gradually rises as shown by many X-ray
data (e.g., Snowden et al. 2008; Cavagnolo et al. 2009; Pratt et al. 2010;
Walker et al. 2012), and as expected on basic astrophysical grounds.

\begin{figure*}
\epsscale{0.8}\plotone{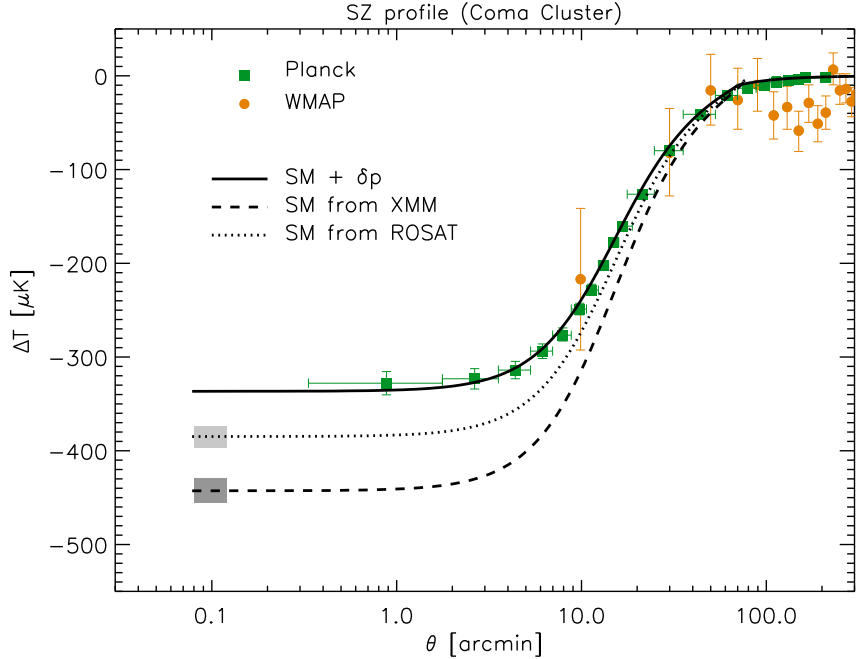}\caption{Profile of SZ effect toward the Coma
cluster. Green squares refer to the \textsl{Planck} data by Ade et al.
(2012), and orange circles to the \textsl{WMAP} data by Komatsu et al.
(2011). Dashed line illustrates the SM outcome (smoothed on the
\textsl{Planck} resolution scale) based on the fit to the X-ray data from
\textsl{XMM-Newton} (see \S~2), with the heavy shaded area representing the
associated $2-\sigma$ uncertainty; dotted line and light shaded area
illustrate the same when basing on the fit to the X-ray brightness from
\textsl{ROSAT} data. The solid line is our outcome when a non-thermal
contribution $\delta p/p\approx 20\%$ (or $15\%$) to the pressure is included
in the SM (see Eqs. 2 and 3).}
\end{figure*}

\section{The Supermodel view}

In fact, the spherically-averaged entropy profile is to rise from a central
level $k_c$ into an outer ramp with slope $a$ toward the virial boundary $R$,
following the pattern (Voit 2005; Lapi et al. 2005)
\begin{equation}
k(r)=k_c + k_R\, (r/R)^a~.
\end{equation}
Specifically, a central baseline $k_c\sim 10^2$ keV cm$^2$ is produced during
the early collapse and virialization of the cluster core; then the
intergalactic gas is condensed to levels $n\sim 10^{-3}$ cm$^{-3}$ in step
with the general overdensities around 200 over the average background, while
it is heated up to temperatures $k_B T\approx G\, M/10\,R \sim$ a few keVs.
These conditions imply thermal pressures $p\approx 2\,n\, k_B T$, of order a
few $10^{-11}$ erg cm$^{-3}$. The baseline $k_c$ may be subsequently lowered
by radiative cooling, or enhanced by AGN feedback (see Lapi et al. 2003;
Fabian 2012) and deep, energetic mergers (see Markevitch \& Vikhlinin 2007;
McCarthy et al. 2007). These processes, respectively, produce current
conditions of the cool core kind with $k_c\sim 10$ keV cm$^2$, or of the
non-cool core kind with $k_c > 10^{2}$ keV cm$^2$ like in Coma (Cavaliere et
al. 2011).

At the outer end, the slope $a\approx 1$ and the boundary value $k_R\sim$
several $10^3$ keV cm$^2$ are originated as entropy is continuously produced
by strong virial accretion shocks (for observational evidence in Coma, see
Brown \& Rudnick 2011; Markevitch 2012), and then is conserved and stratified
as the infalling gas is compressed into the gravitational potential well
(Tozzi \& Norman 2001; Cavaliere et al. 2011).

The SM formalism simply consists in inserting the entropy run of Eq.~(1) in
the differential equation for hydrostatic equilibrium of the ICP (see
Cavaliere et al. 2009 for details); this is easily integrated to obtain
\emph{linked} runs of the thermal pressure
\begin{equation}
{p(r)\over p_R}=\left[1+{2\, G\, m_p\over 5\, p_R^{2/5}}\,\int_r^R{\mathrm
d}x~{M(<x)\over x^2\, k^{3/5}(x)}~\right]^{5/2}~,
\end{equation}
of the density $n(r)\propto [p(r)/k(r)]^{3/5}$, and of the temperature
$T(r)\propto p^{2/5}(r)\, k^{3/5}(r)$. Here $p_R$ is the value at the virial
boundary, while $M(<r)$ is the gravitational mass distribution mainly
contributed by the dark matter (see Lapi \& Cavaliere 2009).

In Fig.~1 we show how the SM fits the projected profiles of X-ray brightness
$S_X\propto n^2\, T^{1/2}$ found by Churazov et al. (2012), and of the
emission-weighted temperature $T$ measured by Snowden et al. (2008) and Wik
et al. (2009) in Coma; the reader is referred to Fusco-Femiano et al. (2009)
for details. From these fits we extract values (with their $1-\sigma$
uncertainty, and consistent with the last reference) of the parameters
$k_c\approx 535\pm 180$ keV cm$^2$, $a\approx 1.3\pm 0.2$, and $k_R\approx
5050\pm 225$ keV cm$^2$ specifying the entropy pattern in Eq.~(1).

Based on such values, Coma turns out to be an \emph{extreme} HE (high
entropy) cluster both in the \textit{core} and in the \textit{outskirts},
according to the classification used in Cavaliere et al. (2011). As discussed
there, such conditions call for impacts of several energetic mergers down to
the center that deposit energies of order $10^{64}$ ergs, and for strong
accretion shocks standing at the virial boundary with Mach numbers
$\mathcal{M}\sim 10$ that produce outer temperatures $k_B T_R\sim 5$ keV.
Both processes are in tune with the rich, supercluster environment that
surrounds Coma.

From Eq.~(2) we find the radial pressure profile $p(r)$ with no recourse to
delicate deprojections.  Thence we obtain the \emph{thermal} electron
pressure $p_e$, and compute the profile of the corresponding SZ
Comptonization parameter $y$ (cf. \S~1). In Fig.~2 we express our result in
terms of the equivalent Rayleigh-Jeans decrement $\Delta T\equiv -2\, y\,
T_{\rm CMB}$ of the CMB temperature $T_{\rm CMB}\approx 2.73$ K, to compare
with the \textsl{Planck} measurements as presented by Ade et al. (2012), see
their Fig.~4.

\begin{figure*}
\epsscale{0.8}\plotone{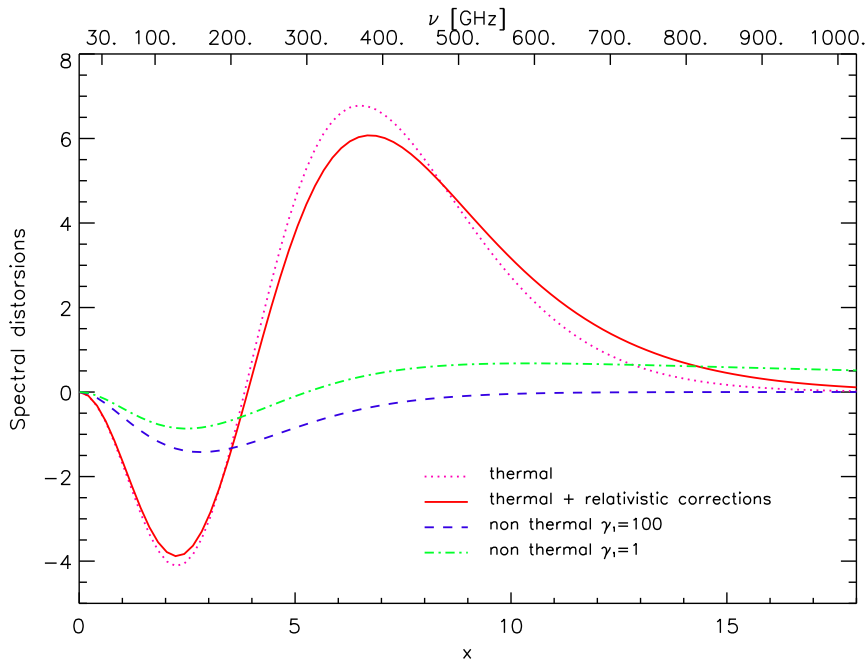}\caption{Full spectral distorsions of the CMB
intensity due to the SZ effect, computed for the Coma cluster. The lower
scale represents the quantity $x\equiv h\nu/k_B T_{\rm CMB}$, while the upper
scale is labeled with the frequency $\nu$ in GHz; note that \textsl{Planck}
is sensitive to bands in the range $\nu\approx 100-857$ GHz. Magenta dotted
line refers to the thermal SZ effect; red solid line is the thermal effect
with the relativistic corrections for the Coma average temperature $k_B T =
8.2$ keV; blue dashed line is the non-thermal SZ effect from relativistic
electrons down to $\gamma_1=10^2$ with density $10^{-7}$ cm$^{-3}$, and green
dot-dashed line down to $\gamma_1=1$ with density $10^{-5}$ cm$^{-3}$. Note
in the latter the high frequency tail, and the displacement of the null from
the thermal value, see \S~4 for details. A powerlaw electron energy
distribution with spectral index $s=3.4$ is adopted.}
\end{figure*}

\section{Comparing the SZ and X-ray views of Coma}

Fig.~2 highlights a \emph{deficit} in the values of $|\Delta T|$ as measured
by \textsl{Planck}, relative to those expected from the X-ray pressure. The
discrepancy appears to be remarkably sharp at the center, well beyond the
uncertainties budget presented by Ade et al. (2012).

Such a SZ vs. X-ray mismatch goes also beyond the uncertainties affecting the
entropy parameters from the X-ray fits (cf. shaded areas in Figs.~1 and 2),
with effects damped out by the weak dependence on $k(r)$ of the integral term
in Eq.~(2). The mismatch may be marginally alleviated if one relied on the
X-ray data from \textsl{ROSAT} instead of \textsl{XMM-Newton} that with its
higher-resolution instruments may enhance the clumpiness effects, so biasing
high the brightness (see Churazov et al. 2012). On the other hand, we recall
from \S~1 that an analogous SZ vs. X-ray mismatch is obtained with quite
different fitting tools by Ade et al. (2012). Thus the tension turns out to
be \emph{model-independent}, and calls for a physical explanation in terms of
a pressure contribution adding to the thermal value $p$.

In this context, the SM formalism is endowed with an extra gear (Cavaliere et
al. 2011), i.e., its ability to straightforwardly include in the equilibrium
a \textit{non-thermal} component $\delta p$ to yield the total pressure
$p+\delta p$. The result can be simply described in terms of Eq.~(2), with
$p$ and $k$ rescaled to
\begin{equation}
\hat{p}\equiv p\,(1+\delta p/p)~~~~~,~~~~~\hat{k}\equiv k\,(1+\delta p/p)~.
\end{equation}
In the above paper we have discussed in one particular instance (focused on
turbulence in cluster outskirts) how the pressure $\delta p(r)$ can be
physically characterized in terms of a normalization provided by the infall
kinetic energy seeping through the virial shocks to drive turbulence, and of
a dissipative decay scale.

For Coma a decay scale is not needed, and a nearly uniform $\delta p/p$
applies to a good approximation. Thus the net outcome is to \emph{lower} the
normalization applying to the thermal pressure at the virial radius, to read
$p_R\propto (1+\delta p/p)^{-1}$. Resolving the tension between the SZ vs.
the X-ray data requires $\delta p/p\approx 15\%$ (up to $20\%$ for
\textsl{XMM-Newton}, which may however include a $5\%$ bias due to
clumpiness, see above). The outcome is illustrated in Fig.~2 by the solid
line; we remark that while the SZ profile from the SM has not been derived
from a formal fit, yet it turns out to represent well the \textsl{Planck}
data over their whole radial range. In particular, the \textit{thermal}
pressure derived with the SM is now lower by $\approx 15\%$, as in fact
sensed by the SZ effect.

Note that a uniform $\delta p/p$ implies the density $n\propto (p/k)^{3/5}$
to be closely unaltered, while the temperature normalization is affected as
$T_R\propto p_R/n_R\propto (1+\delta p/p)^{-1}$; this amounts to a minor
recalibration in the strength of the virial shocks (see \S~2; also Cavaliere
et al. 2009; 2011). The resulting temperature profile stays close to that in
Fig.~1 (bottom panel) with the thermal component of the central entropy
recalibrated to $k_c\approx 470$ from the previous value around $540$ keV
cm$^2$.

To sum up, the thermal electron pressure is related to the equilibrium
pressure $\hat{p}$ by
\begin{equation}
p_e\approx {0.52\,\hat{p}\over 1+\delta p/p}~.
\end{equation}
With $\delta p/p\approx 15-20\%$, this boils down to $p_e\approx 0.45-0.42\,
\hat{p}$, definitely \emph{lower} than the bound $0.5\, p$ pointed out in
\S~1. Note that sensible variations in the ICP metallicity $Z\approx 0.4\pm
0.03$ measured in Coma by Sato et al. (2011) would bias only by a few
percents the electron pressure inferred from the X-ray bremsstrahlung
radiation, as discussed by Churazov et al. (2012).

\section{Discussion and conclusions}

Next we discuss the physical nature of such a \textit{non-thermal }pressure
contribution $\delta p$ to the overall equilibrium.

\begin{itemize}

\item Cosmic-ray protons potentially constitute attractive contributors
    (e.g., Pfrommer et al. 2005), as their energy is longlived and can be
    stored within a cluster. However, in Coma their overall energy
    density has been bounded to be less than a few $10^{-2}$ of the
    thermal pressure by radio and $\gamma$-ray observations (see
    Ackermann et al. 2010; Bonafede et al. 2012). On the other hand,
    cosmic rays may play a role as injectors of secondary electrons, to
    be subsequently accelerated by turbulence and shocks in the ICP (see
    Brunetti et al. 2012).

\item Ongoing turbulence originated by recent mergers that drive plasma
    instabilities in the weakly magnetized ICP constitutes an attractive
    contributor in view of its direct link to the primary energetics.
    Such a turbulence has been discussed by many authors as a source of
    velocity and density fluctuations (see Nagai et al. 2007; Vazza et
    al. 2010; Iapichino et al. 2011); it is widely held to accelerate
    with moderate efficiency supra-thermal electrons in the plasma to
    mildly relativistic energies giving rise to steep distributions (see
    Schlickeiser et al. 1987; Sarazin \& Kempner 2000; Blasi et al. 2007;
    Brunetti \& Lazarian 2011). However, in Coma the density fluctuations
    caused by ongoing subsonic turbulence have been constrained by
    Churazov et al. (2012; see their \S~5.2 and 5.3) to be less than
    $5\%$ on scales $30-300$ kpc. The corresponding indirect estimates of
    current turbulent velocities $\la 450$ km s$^{-1}$ would fall short
    of providing the additional pressure required to relieve the SZ vs.
    X-ray tension. The actual turbulence velocities will be directly
    probed with the upcoming \textsl{ASTRO-H} mission
    (\texttt{http://www.astro-h.isas.jaxa.jp/}).

\item Relativistic electrons with Lorentz factors $\gamma\ga 10^3$ in the
    diffuse magnetic field $B\approx$ a few $\mu$G measured in Coma emit
    the large-scale synchrotron radiation observed at $\nu\ga 30$ MHz in
    the form of the classic Coma radiohalo, see Govoni et al. (2001) and
    Brunetti et al. (2012). Based on the halo shape discussed in the last
    reference, the pressures of the magnetic field and of the energetic
    electrons appear to be effectively coupled to that of the dominant
    thermal population (see discussions by Brown \& Rudnick 2011 and
    Bonafede et al. 2012). The integrated radio power of several
    $10^{40}$ erg s$^{-1}$ implies a relativistic energy density of order
    $10^{-16}$ erg cm$^{-3}$ (see Giovannini et al. 1993, with parameters
    updated). Although the corresponding pressure value is substantially
    smaller than the required $\delta p\approx 0.15\, p\approx$ several
    $10^{-12}$ erg cm$^{-3}$, relativistic electrons can provide
    interesting candidates if their energy distribution extends steeply
    toward a lower end $\gamma_1\lesssim 10^2$.

\end{itemize}

Such an extension is consistent with the radio spectrum retaining a slope
$\alpha\approx 1.2$ or somewhat steeper, as observed down to frequencies
$\nu\approx 31$ MHz (see Henning 1989); the corresponding electron
distribution is to rise toward low energies as $\gamma^{-s}$ with slope
$s\equiv 2\alpha+1\approx 3.4$. Existing data (cf. Henning 1989) also show
that at lower frequencies the radio flux in Coma is still sustained, and may
even feature a steeper component, as found in other clusters (see van Weeren
et al. 2012); \textsl{LOFAR} will soon clear the issue (see
\texttt{http:/www.lofar.org/}).

The amount of \emph{non-thermal} pressure implied by the above electron
population may be estimated as $\delta p\approx \gamma_1\,m_e c^2\, n_{\rm
rel}(\gamma_1)/3\propto \gamma_1^{2-s}$, and refined with the full
expressions including mildly relativistic electrons as given by En{\ss}lin \&
Kaiser (2000, their Appendix A). Using the value $2\times 10^{40}$ erg
s$^{-1}$ of the radiohalo luminosity at $100$ MHz and the profile given by
Brunetti et al. (2012), we compute that a non-thermal contribution $\delta
p/p\approx 15\%$ would indeed obtain on extending a straight electron
distribution down to $\gamma_1\sim$ a few.

On the other hand, a slope sustained against the fast Coulomb losses (e.g.,
Sarazin 1999; Petrosian \& East 2008) requires such electrons cannot be drawn
from the thermal pool, but rather to have been injected over a few $10^7$ yr
by the action of mergers, or by AGNs (like the current sources associated
with NGC 4869 and NGC 4874), or by cosmic-ray interactions (see Brunetti et
al. 2012). These electrons are widely held to be accelerated via turbulence
and low-$\mathcal{M}$ shocks, recently driven by mergers deep and energetic
but already on the way of dissipating, so as to meet the constraints set by
Churazov et al. (2012) and recalled above. We have stressed at the end of
\S~2 that similar merging events over timescales of Gyrs are
\emph{independently} required for providing the top level $k_c\approx 500$
keV cm$^2$ of the central entropy in Coma.

Energy distributions steep down to $\gamma_1 \sim$ a few imply a density
$n\sim 10^{-5}$ cm$^{-3}$ in trans-relativistic electrons, and so provide an
upper bound for gauging the actual low-$\gamma$ electron population via the
tail of the SZ effect at very high frequencies $\ga 1$ THz, and the
displacement of the thermal null at $217$ GHz; these spectral distorsions are
illustrated in Fig.~3 (see also Rephaeli 1995). Such features in the SZ
spectrum are within the reach of sensitive instrumentations like
\textsl{ALMA} (see \texttt{http://www.almaobservatory.org/}).

In the range $\gamma < 10^2$ the electron distribution will be progressively
flattened down by Coulomb interactions over timescales $< 10^{-1}$ Gyr. But a
`silent pool' of cooling electrons with $\gamma \sim 10^2$ can be replenished
and piled up since their lifetimes top at about $1$ Gyr. With a cumulative
density $n\sim 10^{-7}$ cm$^{-3}$ resulting from many mergers, these
electrons can yield a non-thermal contribution $\delta p/p\approx 15\%$.
Their synchrotron and relativistic bremsstrahlung radiations would easily
escape detection (Sarazin 1999; Sarazin \& Kempner 2000), while their
collective contribution to pressure is \emph{probed} just by the thermal SZ
effect. Note that to sustain such a silent pool would require a rather high
energy dissipation rate by mergers.

In summary, the intriguing physical conditions featured by the inner ICP in
the Coma cluster include both a \emph{thermal} and a \emph{non-thermal}
component, to be probed via three observational channels across the
electromagnetic spectrum: the bremsstrahlung emission in X rays, the thermal
and relativistic SZ effects in microwaves, and the diffuse synchrotron
radiation in the radio band. On comparing the first two views, we have found
a model-independent \emph{tension} between the SZ \textsl{Planck} data and
the X-ray observations. In fact, a similar mismatch has been also found by
Ade et al. (2012) with their empirical fitting formulae, concerning not only
spherically-averaged profile, but also $3$ out of $4$ angular sectors probed
in detail.

On large scales, the tension is hard to explain out in terms of overall
asphericities given their limited impact in Coma, as discussed by De Filippis
et al. (2005). On small scales, the narrow shock jumps reported in selected
sectors by \textsl{Planck} (Ade et al. 2012) are diluted by line-of-sight
projection and azimuthal averaging. On intermediate scales of order several
$10^2$ kpc, the presence in the ICP of substructures (see Ade et al. 2012)
and fluctuations (see Khedekar et al. 2012) may contribute to locally bias
the X-ray pressure; however, given the constraints on the density
fluctuations in Coma (Churazov et al. 2012), we expect these effects to be
limited to about $5\%$.

We have instead proposed that the SZ vs. X-ray tension can be resolved in
terms of a physical condition, i.e., \emph{non-thermal} support $\delta
p/p\sim 15-20\%$ yielding a lower thermal electron pressure $p_e\approx
0.52\,\hat{p}/(1+\delta p/p)\approx 0.45-0.42\, \hat{p}$, see Eq.~(4). The
spherical SM proceeds to  provide a pressure profile that straightforwardly
incorporates  the \textit{non-thermal} contribution to pressure, and fits in
detail the SZ shape while being \textit{consistent} with the resolved X-ray
observables (see Figs.~1 and 2). We stress that, at variance with the fitting
formulae used by Ade et al. (2012), the SM pressure profile also features a
slow decline for $r\gtrsim 0.3\, R$ in agreement with the \textsl{Planck} SZ
data. Related features are also pleasingly apparent in Fig.~1 of Lapi et al.
(2012), where the SM is found to perform considerably better than the fitting
formulae by Arnaud et al. (2010) in reproducing the stacked SZ profile from
several clusters observed with the \textsl{South Pole Telescope}. These
successes support our view that the discrepancy between X rays and SZ effect
in Coma is not dominated by specific asymmetries.

Given the current constraints on cosmic-rays and on turbulence (pending
direct measurements of the turbulent velocities), we have discussed the
additional non-thermal pressure in terms of a mildly relativistic electron
population with $\gamma$ in the range from a few to $10^2$. We stress such a
component to constitute a natural \emph{byproduct} of the intense merger
activity independently required for yielding the very \emph{high} central
entropy in the thermal intracluster plasma of the Coma cluster.

\setcounter{footnote}{0}

\begin{acknowledgements}
We thank our referee for constructive comments. Work supported in part by
ASI, INAF, and MIUR. We thank E. Churazov, J. Gonzalez-Nuevo, P. Mazzotta,
and P. Natoli for useful discussions. A.L. thanks SISSA for warm hospitality.
\end{acknowledgements}

\end{document}